\documentclass[superscriptaddress,showpacs,preprintnumbers,amsmath,amssymb,twocolumn,prb]{revtex4-1}
\usepackage{graphicx}
\usepackage{verbatim} 
\usepackage[dvips]{color} 
\usepackage{epstopdf}

\usepackage{dcolumn}
\usepackage{bm}
\usepackage{upgreek}
\usepackage{pdfpages}

\begin{document}

\preprint{APS/123-QED}

\title{Photon echoes from (In,Ga)As quantum dots embedded in a Tamm-plasmon microcavity}

\author{M.~Salewski}
 	\email{matthias.salewski@tu-dortmund.de}
 	\affiliation{Experimentelle Physik 2, Technische Universit\"at Dortmund, 44221 Dortmund, Germany}
\author{S.~V.~Poltavtsev}
 	\affiliation{Experimentelle Physik 2, Technische Universit\"at Dortmund, 44221 Dortmund, Germany}
 	\affiliation{Spin Optics Laboratory, St. Petersburg State University, St. Petersburg 198504, Russia}
\author{Yu.~V.~Kapitonov}
 	\affiliation{St. Petersburg State University, St. Petersburg 199034, Russia}
\author{J.~Vondran}
 	\affiliation{Experimentelle Physik 2, Technische Universit\"at Dortmund, 44221 Dortmund, Germany}
\author{D.~R.~Yakovlev}
 	\affiliation{Experimentelle Physik 2, Technische Universit\"at Dortmund, 44221 Dortmund, Germany}
 	\affiliation{Ioffe Physical-Technical Institute, Russian Academy of Sciences, 194021 St. Petersburg, Russia}
\author{C.~Schneider}
 	\affiliation{Technische Physik, Universit\"at W\"urzburg, 97074 W\"urzburg, Germany}
\author{M.~Kamp}
 	\affiliation{Technische Physik, Universit\"at W\"urzburg, 97074 W\"urzburg, Germany}
\author{S.~H\"ofling}
 	\affiliation{Technische Physik, Universit\"at W\"urzburg, 97074 W\"urzburg, Germany}
\author{R. Oulton}
    \affiliation{Quantum Engineering Technology Labs, H. H. Wills Physics Laboratory and Department of Electrical \& Electronic Engineering, University of Bristol, BS8 1FD, UK}
\author{I.~A.~Akimov}
 	\affiliation{Experimentelle Physik 2, Technische Universit\"at Dortmund, 44221 Dortmund, Germany}
 	\affiliation{Ioffe Physical-Technical Institute, Russian Academy of Sciences, 194021 St. Petersburg, Russia}
\author{A.~V.~Kavokin}
 	\affiliation{Spin Optics Laboratory, St. Petersburg State University, St. Petersburg 198504, Russia}
	\affiliation{School of Physics and Astronomy, University of Southampton, SO17 1 BJ, Southampton, United Kingdom}
	\affiliation{CNR-SPIN, Viale del Politecnico 1, I-00133 Rome, Italy}
\author{M.~Bayer}
 	\affiliation{Experimentelle Physik 2, Technische Universit\"at Dortmund, 44221 Dortmund, Germany}
 	\affiliation{Ioffe Physical-Technical Institute, Russian Academy of Sciences, 194021 St. Petersburg, Russia}

\date{\today}


\begin{abstract}
We report on the coherent optical response from an ensemble of (In,Ga)As quantum dots (QDs) embedded in a planar Tamm-plasmon microcavity with a quality factor of approx. 100. Significant enhancement of the light-matter interaction is demonstrated under selective laser excitation of those quantum dots which are in resonance with the cavity mode. The enhancement is manifested through Rabi oscillations of the photon echo, demonstrating coherent control of excitons with picosecond pulses at intensity levels more than an order of magnitude smaller as compared with bare quantum dots. The decay of the photon echo transients is weakly changed by the resonator indicating a small decrease of the coherence time $T_2$ which we attribute to the interaction with the electron plasma in the metal layer located close (40~nm) to the QD layer. Simultaneously we see a reduction of the population lifetime $T_1$, inferred from the stimulated photon echo, due to an enhancement of the spontaneous emission by a factor of 2, which is attributed to the Purcell effect, while non-radiative processes are negligible as confirmed from time-resolved photoluminescence.
\end{abstract}


\maketitle


Light-matter interaction in photonic nanostructures attracts strong attention in all areas of optics. Efficient coupling at the nanoscale plays a decisive role for realization of single photon emitters and other nonclassical light sources of importance in quantum information technologies.~\cite{photonic-Obzor, Langbein-2013} Various structural concepts based on photonic crystals, patterned microcavities or plasmonic structures have been intensely studied in that respect.\cite{System-types} Another interesting system is a Tamm-plasmon (TP) resonator in which confinement of the optical field is obtained between a distributed Bragg reflector (DBR) and a thin metal layer, leading to the appearance of a TP photonic mode.\cite{Kaliteevski-07} In addition, TP structures support surface plasmon polaritons (SPPs), evanescent electromagnetic waves at the metal-semiconductor interface which can propagate along this interface. Especially interesting in such a system is the coupling between SPPs and TP cavity modes.\cite{Fedyanin-SPP13,Oulton-SPP14,Bellessa2016} Therefore these structures are appealing for generation of SPPs via optical or electrical pumping of a close active layer containing, e.g., semiconductor quantum dots (QDs). In general, the integration of semiconductors into plasmonic circuits is appealing for compensating losses or switching in these circuits. Furthermore, the metal mirror may be used as an electrode to apply a bias voltage for controlling the charging state of the QDs or pumping the optically active layer electrically.\cite{Kamp-14} As active material in the resonator single or multiple quantum well (QW)~\cite{Baumberg-11,Bellessa-13,Kamp-14,Hommel-15} or QD~\cite{Senellart-PRL11,Kamp15} layers as well as organic materials~\cite{Leo-organics2012,Chen-organics2014} were used. So far, efforts have mainly been focused on time-integrated and time-resolved studies of the emission under non-resonant excitation. Thereby the Purcell effect of single QD excitons coupled to a localized TP mode,\cite{Senellart-PRL11} enhancement of the spontaneous emission~\cite{Kamp15} and coherent laser emission~\cite{Bellessa-13,Leo-organics2012} were demonstrated.

Coherent spectroscopy such as transient four-wave mixing (TFWM) is a powerful tool for investigating the nonlinear optical phenomena and coherent dynamics of optical excitations confined in semiconductor nanostructures.\cite{Shah-book} It allows one to study ensembles of light emitters and to perform direct measurements of their times for dephasing, decoherence, and population decay. Moreover, in the strong field regime Rabi oscillations can be used for direct evaluation of the light-matter interaction strength. TFWM and two-dimensional Fourier spectroscopy were used to investigate the coherent optical response of exciton-polaritons in QW and QD based microcavities with high quality factors (strong coupling regime).\cite{Langbein-2013, Deveaud-12, Bristow-15} QDs were also implemented in low quality DBR-based cavities in order to increase the strength of TFWM signal and to study the corresponding coherent optical phenomena.\cite{Fras-2016,Poltavtsev2016,Mermillod-2016} However, to the best of our knowledge there are no studies of the nonlinear optical phenomena under resonant excitation in TP resonators.

\begin{figure*}[]
\includegraphics[width=17cm]{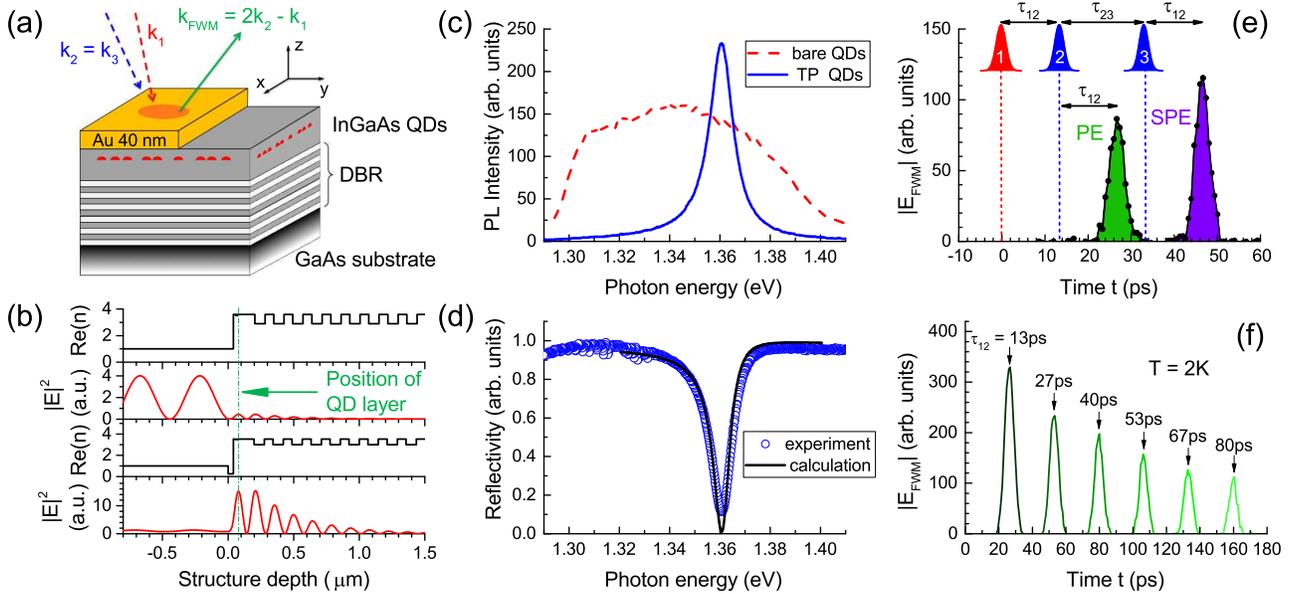}
\caption{(Color online) (a) Schematic presentation of the investigated TP structure and the TFWM geometry. (b) Refractive index and electric field distribution under normal incidence of light (along the $z$-axis) for the uncovered part (top) and the metal-covered part (bottom) of the sample. In the TP microcavity the intensity of light at the QD layer position is increased by a factor of 34 compared to the bare waver. (c) Photoluminescence spectra of the inhomogeneously broadened QD ensemble with and without TP microcavity. Photon excitation energy $\hbar\omega_{\rm exc}=2.33$~eV, temperature $T=2$~K. (d) Measured and calculated reflectivity spectra of the TP microcavity. Quality factor $Q=130$, temperature $T=8$~K. The calculation used a transfer matrix method with $d_{w}=118$~nm.\cite{supplement}  (e) and (f) Photon echoes from the TP cavity with pulse energies $\mathcal{P}_1=0.08$~nJ and $\mathcal{P}_2 \approx \mathcal{P}_3 \approx 0.3$~nJ, temperature $T=2$~K. (e) TFWM amplitudes measured at fixed delay times of $\tau_{12}=13\,$ps and $\tau_{23}=20\,$ps. The photon echo (PE) and stimulated photon echo (SPE) signals appear at $t=2\tau_{12}$ and $t=2\tau_{12}+\tau_{23}$, respectively. (f) PEs for different time delays $\tau_{12}$ indicated at the arrows.}
\label{fig:R-PL}
\end{figure*}

The unique feature of TP structures is the possibility to use an arbitrary in-plane design of a metallic layer on top of the semiconductor. Covering only a part of the sample with a gold film allows comparison of the strength of light-matter interaction within the same QD ensemble so that size distribution and QD density remain the same. Here, we show that the coherent optical response of planar QD-TP structures to picosecond laser pulses occurs in form of photon echoes (PEs): (i) The magnitude of the PE signal and its dependence on excitation intensity differ drastically from bare QDs. From Rabi oscillations we estimate the enhancement factor for the driving optical field in the TP structure. (ii) From two-pulse and three-pulse PE transients we evaluate the decoherence and population decay times of excitons in the TP microcavity and compare them with bare QDs.


The investigated structures and experimental approach are summarized in Fig.~\ref{fig:R-PL}. The sample was grown by molecular-beam epitaxy on a GaAs (001) semi-insulating substrate. It comprises 20 pairs of $\lambda/4n$ GaAs/AlAs layers forming the back DBR mirror with the stop-band located in the spectral region of interest ($\sim 1.2-1.4$~eV). A single (In,Ga)As QD layer in the GaAs cavity layer is located about 125~nm above the DBR and 40 nm below the surface. The thickness of the GaAs layer $d_w$ beneath the QDs is slightly varied by a wedge design which allows tuning the cavity resonance by the position of the excitation spot along the gradient axis $x$ (see Fig.~\ref{fig:R-PL}(a)). Half of the sample is covered with a 40 nm thick gold layer, which leads to formation of the TP photon mode.

The QD layer is located at the maximum of the electric field distribution, $40\,$nm away from the gold layer (see Fig.~\ref{fig:R-PL}(b)). It is close enough to the gold to ensure that the coupling to the TP mode is strong, but far enough so that quenching of the photoluminescence (PL) due to surface states, in particular at the uncovered surface, does not occur. The QD density is about $2 \times 10^9\,$cm$^{-2}$, their height and lateral sizes before overgrowth are about 2.3~nm and 25~nm, respectively.\cite{growth} In order to prevent tunneling of photoexcited carriers into the metal layer, a 10~nm Al$_{0.2}$Ga$_{0.8}$As barrier was introduced between the QDs and the surface (20~nm below the surface). We also studied structures with a single layer of (In,Ga)As QDs which is embedded in a GaAs $\lambda$ microcavity of similar quality factor formed by two DBR mirrors.\cite{Poltavtsev2016}

The low temperature ($T=2$~K) photoluminescence spectrum from the bare QDs under non-resonant excitation with photon energy $\hbar\omega_{\rm exc} = 2.33$~eV shows a broad spectral band centered at about 1.35~eV (Fig.~\ref{fig:R-PL}(c)). The large inhomogeneous broadening of $\sim 100$~meV originates from fluctuations of QD size and composition in the ensemble. The emission from the same part of the sample, but covered with gold shows a resonant enhancement of the PL signal around the photon energy of the cavity mode. The PL has a significantly narrower bandwidth with a full width at half maximum (FWHM) of only 12~meV. The maximum of the PL peak and its width correspond to the dip in the cavity reflectivity spectrum (Fig.~\ref{fig:R-PL}(d)). The latter is attributed to a TP photonic mode with energy $\hbar\omega_{\rm TP}=1.36$~eV and width $\Delta=10.5$~meV, corresponding to a quality factor of $Q=130$. Calculations using the transfer matrix method are in good agreement with the measured reflectivity spectrum for $d_w= 118$~nm. The expected enhancement factor of the light intensity at the position of the QD layer is 34 (see Fig.~\ref{fig:R-PL}(b)).\cite{supplement}

The coherent optical response is measured using degenerate TFWM with a sequence of three spectrally narrow ps-pulses in non-collinear reflection geometry as shown in Fig.~\ref{fig:R-PL}(a). A mode-locked tunable Ti:Sa laser with a repetition frequency of 75.75~MHz was used as source of the optical pulses. Pulse 1 with wavevector ${\bf k_1}$ hits the sample at an incidence angle of $6^\circ$ while the following pulses 2 and 3 hit the same spot at $7^\circ$ (${\bf k_2} = {\bf k_3}$). The beams are focused into a spot with a diameter of approx. 200~$\mu$m. The sample is kept at $T=2$~K. The transients are measured by taking the cross-correlation of the TFWM signal $E_{\rm FWM}(t)$ with the reference pulse using heterodyne detection.\cite{Langbein-book} The spectral width of the optical pulses of 0.8~meV is significantly smaller than the width of the resonator mode, i.e. the lifetime of the TP mode in the microcavity is shorter than the pulse duration of $\tau_d=2$~ps.  At zero magnetic field resonantly excited QDs can be considered as isolated two-level systems. Note that excitation of few-particle complexes such as biexcitons is excluded because the spectral width of the laser line is below the biexciton binding energy of about 2~meV. From these parameters we expect that the TFWM signal is determined by photon echoes due to the inhomogeneous broadening of the optical transitions in the QD ensemble. In the case of the TP microcavity the broadening is given by the spectral width of the photonic mode.~\cite{Poltavtsev2016}

An example of the TFWM signal measured under resonant excitation in the TP mode is shown in Fig.~\ref{fig:R-PL}(e). Here, the delay time between pulses 1 and 2 is set to $\tau_{12}=13$~ps, while the delay time between pulses 2 and 3 is $\tau_{23}=20$~ps. In full accordance with our expectation, we observe spontaneous (PE) and stimulated (SPE) photon echoes which appear exactly at time delays of $t=2\tau_{12}$ and $t=2\tau_{12}+\tau_{23}$, respectively. In addition, Fig.~\ref{fig:R-PL}(f) demonstrates that the PE pulse appears at twice the delay between pulses 1 and 2. In what follows, we consider the maximum value of the PE amplitude at the peaks of the PE, $P_{\rm PE}$, or the SPE, $P_{\rm SPE}$, signals.

\begin{figure}[htb]
\includegraphics[width=0.8\columnwidth]{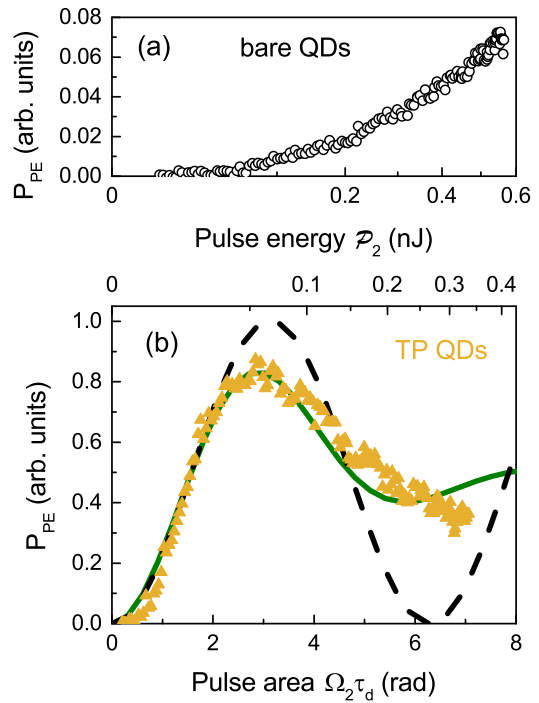}
\caption{(Color online) Dependence of the PE amplitude $P_{\rm PE}$ on the energy of the second pulse $\mathcal{P}_2$ for the bare QDs (a) and the TP microcavity (b). $\mathcal{P}_1=0.026$~nJ, $\tau_{12}=67$~ps, $T=2$~K. The scaling of the $x$-axes (bottom in (a) and top in (b)) is chosen such that it would scale linearly with the square root of $\mathcal{P}_2$. The bottom $x$-axis in (b) corresponds to the area of pulse 2. The dashed curve is a fit using Eq.~\eqref{Rabi}. The solid curve follows from Eq.~\eqref{Rabi} when including the statistical distribution of dipole moments of the two-level systems with a standard deviation of 30\%.}
\label{Fig:Rabi}
\end{figure}

The dependence of the PE amplitude on the intensity of the excitation pulses can be used to compare the strength of light-matter interaction in both systems. For simplicity we consider the dependence of the two-pulse PE on the energy of second pulse, $\mathcal{P}_2$, shown in Figs.~\ref{Fig:Rabi}(a) and \ref{Fig:Rabi}(b) for the bare QDs and the TP microcavity, respectively, for identical experimental conditions. Strong inhomogeneous broadening of optical transitions can lead to a significant modification of the photon echo transients when the energy of the first pulse is scanned.\cite{Poltavtsev2016} First, in the TP microcavity one achieves significantly stronger PE amplitudes for low pulse energies $\mathcal{P}_2 \leq 0.1$~nJ. Second, in the TP microcavity one observes a non-monotonous behavior which resembles the one expected for Rabi oscillations. Both features indicate a considerable enhancement of the light-matter interaction in the TP cavity which allows us to work in the nonlinear regime already at moderate intensities.

For isolated two-level systems in absence of decoherence processes and when optical pulses with rectangular intensity profile are assumed the amplitude of the PE follows the simple relation~\cite{Berman-book}
\begin{equation}
P_{\rm PE} \propto \sin^2 \left( \frac{\Omega_2 \tau_d}{2} \right),
\label{Rabi}
\end{equation}
where $\Omega_2 = 2|E_2 d|/\hbar$ is the Rabi frequency which is determined by the electric field amplitude of the second pulse $E_2$ and the dipole matrix element of the two-level system $d$. In our case $\tau_d$ is fixed while $E_2 \propto \sqrt{\mathcal{P}_2}$ is varied. We use this expression to evaluate the electromagnetic field amplitude inside the microcavity $E_2^{\rm TP-QD}$. This approximation should be valid also in the strong field limit because the lifetime of the cavity mode is the shortest time scale compared to the duration of the excitation pulses and the radiative lifetime in the QDs. For $\mathcal{P}_2 \leq 0.15$~nJ the PE oscillatory behavior in the TP-QD structure is reasonably well reproduced by Eq.~\eqref{Rabi} with the $\pi$-rotation occuring at $\mathcal{P}_2 \approx 0.06$~nJ. However, for larger pulse areas significant damping of the Rabi oscillations takes place. This is mainly due to a statistical distribution of the dipole moments $d$ in the QD ensemble which results in a variation of the pulse area $\Theta = \Omega_2\tau_d$ and blurring of the Rabi oscillations.~\cite{Bori-Rabi} Assuming a Gaussian distribution with a standard deviation $\sigma\Theta$ centered around $\Theta_0$ we obtain $P_{\rm PE} \propto \int_0^{\infty}\sin^2{[\Theta/2]}\exp{[-\frac{(\Theta_0-\Theta)^2}{2\sigma^2\Theta^2}}]d\Theta$. The best fit to the experimental data is obtained for $\sigma =0.3$ as shown by the solid curve in Fig.~\ref{Fig:Rabi}(b). This value is in good agreement with previous results on similar QD ensembles.\cite{Bori-Rabi} In addition other mechanisms such as the interaction with acoustic phonons can lead to damping of the Rabi oscillations. The latter is expected to be strongest for pulses with durations of several ps as used in our experiment.\cite{Kuhn-05, Reiter-14} 

In contrast to the TP microcavity structure, the $P_{\rm PE}$ for the bare wafer follows a quadratic dependence on $\sqrt{\mathcal{P}_2}$ which indicates that the pulse area is significantly below $\pi$ even for pulse energies as large as $\mathcal{P}_2 \approx 0.5$~nJ. Thus a significant enhancement of the light-matter interaction is clearly present in the TP cavity. Direct comparison of the intensities allows us to estimate the enhancement factor of the electromagnetic field for coherent excitation of the QDs. Taking into account that the PE signal depends on both $\mathcal{P}_1$ and $\mathcal{P}_2$, we estimate that the pulse area in Eq.~\eqref{Rabi} is enhanced by a factor of 6, i.e. the amplitude of the electromagnetic field $E_2^{\rm TP} \approx 6 E_2^{\rm b}$. This is in accordance with the calculated increase of the light intensity by a factor of 34 in the TP microcavity as compared with the bare QD system (see Fig.~\ref{fig:R-PL}(b)).

\begin{figure}[htb]
\includegraphics[width=0.7\columnwidth]{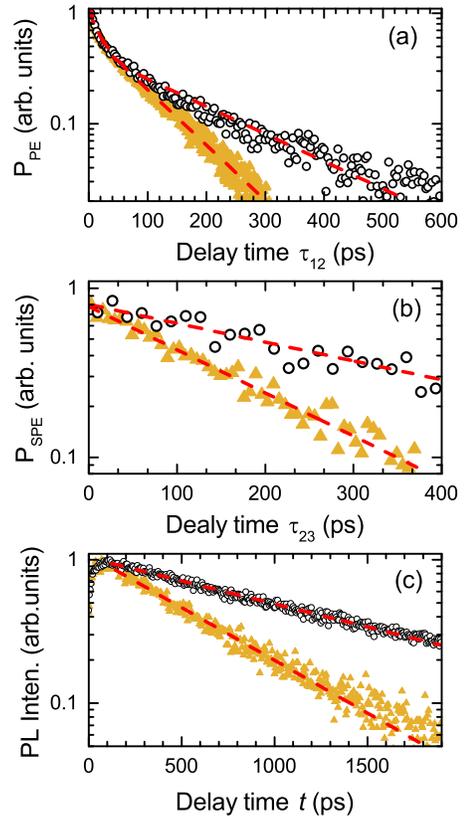}
\caption[width=\textwidth]{(Color online) Normalized time-resolved measurements on the bare QDs (circles) and the QDs in the TP microcavity (triangles). $T=2$~K. (a) PE amplitude as function of the pulse delay $\tau_{12}$. (b) SPE amplitude as function of the pulse delay $\tau_{23}$ in order to measure the lifetime $T_1$. $\tau_{12} = 26.7$~ps. The pulse areas in (a) and (b) did not exceed $\pi$. (c) Time-resolved PL for excitation with photon energy $\hbar\omega_{\rm exc}= 1.494$~eV. The dashed curves are fits with double (a) and single (b,c) exponential decays, respectively. The resulting time constants are summarized in Table~\ref{Tab1}.}
\label{Fig:transients}
\end{figure}

Let us now consider the transient decay of the photon echo signal which gives insight into the coherent dynamics and the relaxation processes in the QD systems. The dependences of the PE and SPE amplitudes on the pulse delays $\tau_{12}$ and $\tau_{23}$ are presented in Figs.~\ref{Fig:transients}(a) and \ref{Fig:transients}(b), respectively. The PE decay reflects an irreversible loss of optical coherence (i.e., of the medium's polarization) due to interaction of the two-level systems with the environment and/or radiative damping. It follows from Fig.~\ref{Fig:transients}(a) that the signal can be described by a double exponential decay $P_{\rm PE} \propto A\exp{(-2\tau_{12}/T_2')} + B\exp{(-2\tau_{12}/T_2)}$ including a short coherence time $T_2'=30$~ps and an approximately 10 times longer coherence time $T_2$. The short dynamics are attributed to fast energy relaxation of QD excitons excited in higher energy states, e.g. in $p$-shell states,\cite{Kurtze-09} into the ground state. On the other hand, the long-lived signal with $T_2$ decay time corresponds to the coherent response of excitons in the ground $s$-shell. Excitation of excitons in different energy shells is possible due to the strong inhomogeneous broadening of the optical transition energies in the studied QDs. Apparently the long component decays faster in the TP microcavity with $T_2^{\rm TP} = 170$~ps, while in the bare QDs $T_2^{\rm b} = 350$~ps, i.e the coherent decay is twice slower (see Table~\ref{Tab1}).

The SPE decay gives insight into the population decay dynamics and allows the evaluation of the exciton lifetime $T_1$. Here, we obtain for the bare QDs $T_1^{\rm b}=390$~ps and for the TP microcavity $T_1^{\rm TP} = 170$~ps. Using $T_2^{-1}=1/2T_1^{-1} + \tau_c^{-1}$ we calculate the pure dephasing time $\tau_c$. $\tau_c$ turns out to be shorter for the excitons confined in the QDs in the TP microcavity, $\tau_c^{\rm TP} = 340$~ps, as compared with the bare QDs, $\tau_c^{\rm b} = 635$~ps. As this time exceeds the exciton lifetime, the pure dephaing is neverheless weak. In a studied fully dielectric DBR structure with $Q \approx 200$ the pure dephasing is negligible and the ground state exciton coherence is radiatively limited: $T_2^{\rm DBR}=2T_1^{\rm DBR}$ with $T_1^{\rm DBR}\approx 300-400$~ps. The latter value has approximately the same magnitude as for bare QDs. From the comparison we conclude that the exciton coherence in the bare QDs is somewhat reduced by charges at the surface 40 nm separated from the dot layer. The gold layer of the TP cavity induces further decoherence which can be attributed to the interaction between excitons and plasmons.

The lifetime measurements deduced from the SPE in Fig.~\ref{Fig:transients}(b) show that for the TP microcavity $T_1$ is approximately halved compared to the bare QDs. This shortening most likely is due to the Purcell effect.\cite{Senellart-PRL11} However, we have to consider also non-radiative processes due to tunneling of photoexcited carriers from the QDs into the nearby metal as potential origin. Further insight can be obtained from PL transients measured using a streak camera for below-barrier pulsed excitation with photon energy $\hbar\omega_{\rm exc} = 1.494$~eV. These measurements are shown in  Fig.~\ref{Fig:transients}(c). The PL signals decay with lifetimes $\tau_0^{\rm b}=1350$~ps and $\tau_0^{\rm TP}=590$~ps for the bare QDs and the TP structure, respectively. Interestingly, the PL decay times $\tau_0$ are significantly larger as compared to the $T_1$-values from the SPE decay. Indeed, the two techniques measure different population dynamics. The PL from the ground state is an incoherent process after relaxation of the involved carriers. While the rise time of this signal is in the few ten ps-range, the PL decay time is significantly extended by several factors. For the chosen conditions a carrier reservoir is excited also in the wetting layer, from where carriers have to be transferred to the QDs. After being captured by the dots, relaxation can occur either by phonon emission or by carrier-carrier scattering. In the latter case a carrier, for example an electron, relaxes at the expense of the other carrier, the hole. Thereby populations in higher states and potentially even again in the wetting layer are created slowing down the carrier recombination. In contrast, SPE is a coherent phenomenon following resonant excitation. If for some reason exciton relaxation occurs, it will not contribute to the echo signal.

\begin{table}[htbp]
\centering
\caption{Decay constants evaluated from Fig.~\ref{Fig:transients}. $T_2$ follows from the PE, $T_1$ from the SPE, and $\tau_0$ from the time resolved PL measurements.}
\begin{tabular}{l c c c c}
\hline\hline
 & $T_1$ (ps) & $T_2$ (ps) & $\tau_c$ (ps) & $\tau_0$ (ps)\\
\hline
bare QDs & $390$ & $350$ & $635$ & $1350$ \\
TP-QDs & $170$ & $170$ & $340$ & $590$ \\
\hline\hline
\end{tabular}\label{Tab1}
\end{table}

Non-radiative processes lead to shortening of the lifetime. PL transients with extended dynamical range allow us to give an upper estimate for the non-radiative rate, $\tau_{\rm NR}^{-1}$, in the TP microcavity which should contribute to the SPE decay rate. The exciton decay rate can be written as $\tau_0^{-1}=\tilde{\tau}_0^{-1} + \tau_{\rm NR}^{-1}$, where $\tilde{\tau}_0^{-1}$ is the radiative decay. Neglecting non-radiative processes in the bare QDs, we set $\tilde{\tau_0}^{\rm b} \approx \tau_0^{\rm b}=1350$~ps. Then from $\tau_0$ = 590~ps, the lower limit for the non-radiative decay time is $\tau_{\rm NR}^{\rm TP} \geq 1$~ns in the TP microcavity. Therefore the shortening of $T_1$ from the 350~ps in the bare QDs to the 170~ps in the TP microcavity cannot be initiated by non-radiative processes, but has to be attributed to the Purcell effect with an enhancement factor of about 2. 

Acceleration of the spontaneous emission rate $T_1^{-1}$ by a factor of 2 represents a significant change of the radiative emission dynamics. In $\lambda/2$ microcavities with ideal planar metal mirrors the Purcell factor is limited to 3, while in structures with DBR mirrors the modification of the spontaneous emission rate is typically smaller than $\pm 20\%$.\cite{Ippen-90,Bayer-01} For planar resonators the spontaneous emission becomes mostly redistributed spatially, while the vacuum field becomes only weakly squeezed leading to a moderate enhancement of the local density of photon modes and the associated emission rate at best. 
In our case, in addition SPPs may become relevant for the shortening of $T_1$. On the other hand, the resonator-induced enhancement of light-matter interaction is much stronger for resonant excitation of coherent processes which is related to the selective excitation of the contributing quantum dots. 

In conclusion, we have demonstrated that the coherent optical response from self-assembled (In,Ga)As QDs embedded in a TP planar microcavity is given by photon echoes. Despite the low quality factor of about 100 we demonstrate a substantial enhancement of the selective optical excitation of QDs whose optical transitions are in resonance with the TP cavity mode. The intensity of the driving optical field is amplified by more than one order of magnitude. Such enhancement allows to observe Rabi oscillations in the photon echo and to perform coherent control of excitons with picosecond optical pulses of moderate intensities, while the statistical distribution of dipole moments still represents a significant problem. The decoherence and population dynamics of excitons in TP structures also experience modifications. We observe a decrease of the radiative recombination time from 350 to 170~ps due to the Purcell effect. The presence of the metal layer gives rise to pure dephasing of the QD excitons with characteristic times of about 200~ps to 400~ps so that pure dephasing remains quite weak. We note that the metal layer at the top of the TP microcavity can be used to control the charge state of QDs electrically.\cite{Kamp15} Therefore, such structures are appealing for investigation of long-lived photon echoes from charged QDs where the decay rate is governed by the spin relaxation of the resident electrons.\cite{NaturePho-memory}

\begin{table*}[]
\centering
\caption{Tamm-plasmon structure composition with layer thicknesses and material parameters used in numerical calculations.}
\begin{tabular}{c c c c}
\hline\hline
 Material & Thickness (nm) & Refractive index $n$ & Function\\
\hline
Au 						&	40	&	$0.17 + 5.6i$ \cite{Johnson-1972}	&	Gold mirror (Top)	\\
GaAs						&	20	&	3.6									&	Cap	\\
Al$_{0.2}$Ga$_{0.8}$As	&	10	&	3.5									&	Electron barrier \\
GaAs						&	10	&	3.6									&	Spacer \\
InAs QDs					&	2.3	&	3.6									&	QDs layer	\\
GaAs						&	10	&	3.6									&	Spacer \\
Si						&	$\sim$0	&	-								&	Si-delta doping	\\
GaAs						&	108 (variable)	&	3.6						&	Mode matching layer \\
$20\times$ AlAs / GaAs		&	78.2 / 66.7 	&	2.9 / 3.6				&	DBR mirror	\\
GaAs						&	thick	&	3.6								&	Substrate (bottom) \\
\hline\hline
\end{tabular}\label{Tab2}
\end{table*}

We are grateful to M.~Glazov and B.~Glavin for useful discussions. We acknowledge the financial support by the Deutsche Forschungsgemeinschaft through the Collaborative Research Centre TRR 142 and the International Collaborative Research Centre 160. S.V.P. and Yu.V.K. thank the Russian Foundation of Basic Research for partial financial support (contracts no. ofi\_m 16-29-03115 and no. 15-52-12016 NNIO\_a). M.B. acknowledges partial financial support from the Russian Ministry of Science and Education (contract no. 14.Z50.31.0021). Yu.V.K. acknowledges Saint Petersburg State University for a research grant 11.42.993.2016. The project ‘SPANGL4Q’ acknowledges financial support from the Future and Emerging Technologies (FET) programme within the Seventh Framework Programme for Research of the European Commission, under FET-Open grant no. FP7-284743.

\section*{Appendix - Sample of Tamm-plasmon structure and simulation method}

\begin{figure}[]
\includegraphics[width=0.7\columnwidth]{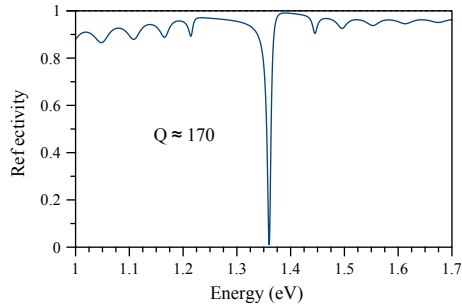}
\caption[width=\textwidth]{(Color online) Reflectivity spectrum simulated for the Tamm-plasmon structure.}
\label{Fig:SOM1}
\end{figure}

\begin{figure}[]
\includegraphics[width=0.7\columnwidth]{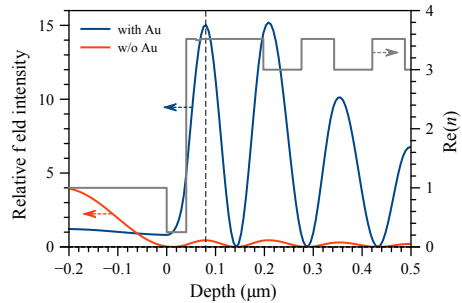}
\caption[width=\textwidth]{(Color online) Relative field intensity distribution inside the structure with top golden layer (blue line) and without it (red line); falling light intensity is 1. Real part of refractive index distributed inside the structure is shown with gray line. The sample top is located at zero; layer above sample is vacuum. Dashed line indicates location of QDs layer.}
\label{Fig:SOM2}
\end{figure}
The composition of the TP structure is presented in Table~\ref{Tab2}. During the growth process substrate was rotated in order to produce homogeneous thicknesses of the layers. The substrate rotation was interrupted for the mode matching layer which resulted in a gradient of thickness of this layer. Owing to this gradient it was possible to tune the spectral position of the photon cavity mode on that part of the sample which was coated with a 40~nm golden layer.

In order to simulate optical properties of the studied TP structure the transfer matrix method was used. In calculation, normal light incidence was considered. The thickness of the mode matching layer was adjusted to get an accordance of the simulated reflection spectrum to the measured one. Resultant spectrum shown in Fig.~\ref{Fig:SOM1} gives quality factor $Q\approx 170$, which is somewhat greater than measured value of 130.

Exactly at the center of the photon mode ($E=1.3607$~eV) the distribution of field intensity inside the structure was calculated for the case of TP structure and for the structure without top golden layer. This calculation is shown in Fig.~\ref{Fig:SOM2} together with the refractive index distribution inside the structure. Enhancement of field intensity at the location of QDs layer is 34.

\end{document}